# Relationship between Magnetic Structure and Ferroelectricity of $LiVCuO_4$

Yukio Yasui, Yutaka Naito, Kenji Sato, Taketo Moyoshi, Masatoshi Sato* and Kazuhisa Kakurai[1]

*Department of Physics, Division of Material Science, Nagoya University,Furo-cho, Chikusa-ku, Nagoya 464-8602*
[1]*Quantum Beam Science Directorate, Japan Atomic Energy Agency,Tokai-mura, Naka-gun, Ibaraki, 319-1195*

Neutron scattering studies and measurements of the dielectric susceptibility ε and ferroelectric polarization $P$ have been carried out in various magnetic fields $\boldsymbol{H}$ for single-crystal samples of the multiferroic system $LiVCuO_4$ with quasi one-dimensional spin 1/2 $Cu^{2+}$ chains formed of edge-sharing $CuO_4$ square planes, and the relationship between the magnetic structure and ferroelectricity has been studied. The ferroelectric polarization is significantly suppressed by the magnetic field $\boldsymbol{H}$ above 2 T applied along $a$ and $b$ axes. The helical magnetic structure with the helical axis parallel to $c$ has been confirmed in $H$=0, and for $\boldsymbol{H}//\boldsymbol{a}$, the spin flop transition takes place at $H$=2 T with increasing $H$, where the helical axis changes to the direction parallel to $\boldsymbol{H}$. The ferroelectric polarization along $a$ at $H$=0 is found to be proportional to the neutron magnetic scattering intensity, indicating that the magnetic order is closely related to the appearance of the ferroelectricity. The relationship between the magnetic structure and ferroelectricity of $LiVCuO_4$ is discussed by considering the existing theories.

Keywords: $LiVCuO_4$, multiferroic, ferroelectricity, helical, magnetic structure

*corresponding author: e43247a@nucc.cc.nagoya-u.ac.jp

Materials which exhibit magnetic and ferroelectric simultaneous transitions are now called multiferroics and attract much attention.[1-5] While almost all systems ever reported have the magnetic moments with spin $S \geq 1$ and/or have more than two magnetic sites, authors' group found that $LiVCuO_4$ is multiferroic system with spins $S$=1/2 and only in the $x^2$-$y^2$ orbital.[6] It is considered to be the first example of spin 1/2 multiferroic systems, which does not bring about any complications due to the multi-orbital and/or multi-site effects in the study of the mechanism of the simultaneous transition.

$LiVCuO_4$ crystallizes to the distorted inverse spinel structure (space group *Imma*),[7] where $Li^+$ and $Cu^{2+}$ ions share the octahedral sites. The one-dimensional (1D) chains of the edge-sharing $CuO_4$ squares or the 1D chains of $Cu^{2+}$ spins ($s$=1/2) are arranged along $b$, which are separated by the nonmagnetic $Li^+$ and $V^{5+}$ ions. In the temperature ($T$) dependence of the magnetic susceptibility, the broad peak attributed to the growth of the short-range spin correlation, typical for low-dimensional antiferromagnets, is observed at ~26 K and the sharp anomaly is observed at the transition temperature $T_N$ =2.4 K, which corresponds to the 3D long-range magnetic order. For this system, the exchange interaction between the nearest-neighbor $Cu^{2+}$ ions, $J_1$ is rather weak (Cu-O-Cu angle is ~95°), or even weaker than the next-nearest-neighbor interaction $J_2$, suggesting that effects of the magnetic frustration are significant in its magnetic properties. Enderle *et al*. reported that $J_1$~-12 K (ferromagnetic) and $J_2$~41 K (antiferromagnetic) by the studies of magnetic excitation using inelastic neutron scattering.[8] Due to the magnetic frustration, the magnetic structure below $T_N$ is expected not to be trivial. Actually, the magnetic structure of $LiVCuO_4$ reported by Gibson *et al*.[9] is helical with the moments in the *ab*-plane (in the $CuO_4$ square planes) and with the incommensurate modulation vector along $b$. Based on the $^7$Li-NMR spectra, the magnetic structure in the applied magnetic field was also discussed.[10] The authors' group reported a clear anomaly at $T_N$ in the $T$-dependence of the dielectric susceptibility ε for the electric field $\boldsymbol{E}//\boldsymbol{a}$, indicating that the ferroelectric transition takes place simultaneously with the magnetic transition.[6]

In the present work, neutron scattering and measurements of dielectric susceptibility ε and ferroelectric polarization $P$ have been carried out on single-crystal samples of $LiVCuO_4$ in the various magnetic fields. Based on results of these studies, the relationship between magnetic structure and ferroelectricity for $LiVCuO_4$ is discussed, in relation to the theoretical models on the occurrence of the multiferroic state.[11-16]

Single-crystal samples of $LiVCuO_4$ used in the measurements of ε and $P$ were grown by a flux method, as reported in the previous paper.[6,17] To avoid the large neutron absorption of Li, we used $^7$Li isotope in the growth of crystals for neutron scattering. First, $^7LiVO_3$ was prepared from the mixtures of $^7Li_2CO_3$ (purity of isotopic enrichment of $^7$Li: 99.94 %) and $V_2O_5$. Polycrystalline samples of $^7LiVCuO_4$ were prepared from the obtained $^7LiVO_3$ and CuO. Then, single-crystal samples of $^7LiCuVO_4$ were grown by a flux method. The obtained crystals were checked not to have appreciable amounts of impurity phases by powder X-ray measurements. The crystal axes were determined by observing the X-ray diffraction lines. The typical size was 20×8×3 mm$^3$.

Neutron scattering measurements were carried out using the triple axis spectrometer HQR(T1-1) of the thermal guide installed at JRR-3 of JAEA in Tokai, where the double axis condition was adopted. The horizontal collimations were 12′(effective)-40′-60′ and the neutron wavelength was 2.4581 Å. The 002 reflection of Pyrolytic graphite (PG) was used as the monochromator. PG filter was placed after the second collimator to suppress the higher-order contamination. In the measurements in $H$=0, the crystal was oriented with the [100] and [010] axes, in one case, and the [010] and [001] ones, in another case, in the scattering plane. The crystal was set in an Al-can filled with exchange gas, and the can was mounted in a liquid $^4$He cryostat. In the measurements under the applied magnetic field, the crystal was oriented with the [100] direction vertical, where the [010] and [001] axes were in the scattering plane. The magnetic field was applied using a superconducting magnet along the vertical direction. In the analyses of the data, the isotropic magnetic form factor for $Cu^{2+}$ were used.[18)]

To study the dielectric susceptibility ε, the capacitance of the thin sample plates was measured, where the electrodes were attached with the silver paint to both sides of the plate surface. An ac capacitance bridge (Andeen Hagerling 2500A) with the frequency of 1 kHz was used. If the stray capacitance is zero, ε is directly proportional to the observed capacitance $C$. The spontaneous polarization $P$ was obtained by measuring the pyroelectric current with an electrometer (Keithley 6517A), where the temperature was swept at a rate of ~0.5 K/min. To avoid the domain formation, we applied an electric field (~1 kV/mm) during the sample-cooling from a temperature well above $T_N$.

Figure 1 shows the $T$-dependence of the capacitance $C$ of $LiVCuO_4$ for $\boldsymbol{E}//\boldsymbol{a}$ taken under various magnetic field. In $H$=0, we have observed a significant anomaly in the $C$-$T$ curves at ~2.4 K, indicating that the ferroelectric transition takes place simultaneously with the magnetic transition. Note that the double-peak structure of the $C$-$T$ curve at $H$=0 shown in Figs. 1(a) and 1(b) seems to be due to an extrinsic origin, because the structure is not reproducible. For $\boldsymbol{E}//\boldsymbol{b}$ and $\boldsymbol{E}//\boldsymbol{c}$, no anomaly in the $C$-$T$ curve was observed (as shown in Fig. 4 of ref. 6), indicating that the ferroelectricity appears only along $a$. We observed the finite spontaneous polarization $P$ along $a$, as shown later.

The behavior of the $C$-$T$ curves is drastically changed by the applied magnetic field. The peak height $\Delta C$ in the $C$-$T$ curves is shown against $H$ in the insets of Figs. 1(a)-1(c). For $\boldsymbol{H}//\boldsymbol{c}$, we can see that $\Delta C$ exhibits a linear decrease and the peak temperature of the $C$-$T$ curve monotonically decreases with increasing $H$. These results are attributed that $T_N$ is suppressed by the magnetic field. For $\boldsymbol{H}//\boldsymbol{a}$ and $\boldsymbol{H}//\boldsymbol{b}$, $\Delta C$ does not exhibit significant changes with $H$ in the region $0 \leq H < 2$ T, but rapidly decreases with further increasing $H$. The peak width of $C$-$T$ curves becomes broader with increasing $H$. These results indicate that some kind of transition at $H$~2 T is induced by the magnetic fields $H$ parallel to $a$ and $b$ or parallel to the $CuO_4$ square planes.

To study of the detailed magnetic structure, we have carried out neutron diffraction measurements on single-crystal samples of $^7LiVCuO_4$. In $H$=0, the neutron scattering intensities have been measured at $\boldsymbol{Q}=(h, k, 0)$ and $(0, k, l)$ in the reciprocal space at 1.6 K (< $T_N$) and 5 K (> $T_N$). At 1.6 K, we have observed the magnetic superlattice reflections at $\boldsymbol{Q}=(h, k\pm\delta, 0)$ and $(0, k\pm\delta, l)$ ($h$ and $l$=odd, $k$=even) with δ~0.466. Examples of the observed ω-scan (sample-angle scan) profiles of 0ζ1 (ζ=0.466) reflection taken at several temperatures are shown in Fig. 2. From the figure, the magnetic ordering is found to grow with decreasing $T$ below $T_N$. The magnetic structure which can reproduce the observed magnetic scattering intensities at 1.6 K is shown schematically in the inset of Fig. 3. It is consistent with the structure reported by Gibson *et al.*[9)] The integrated intensities of the magnetic reflections at 1.6 K are also shown in Fig. 3 against the calculated intensities. (We plot the data of crystals #1 and #2 from different batches.) The details of the obtained structure can be described as follows. The $Cu^{2+}$ moments have a helical structure with the modulation vector of $\zeta\boldsymbol{b}^*$ (or the pitch angle along the 1D Cu-chain is ~ 84 °). The moments at Cu sites shifted by (1/2, 0, 1/2) from a Cu site are antiparallel to that at the original one. The ordered moment at 1.6 K is estimated by comparing the magnetic scattering intensities with the nuclear ones, to be 0.25(±0.03) $\mu_B$.

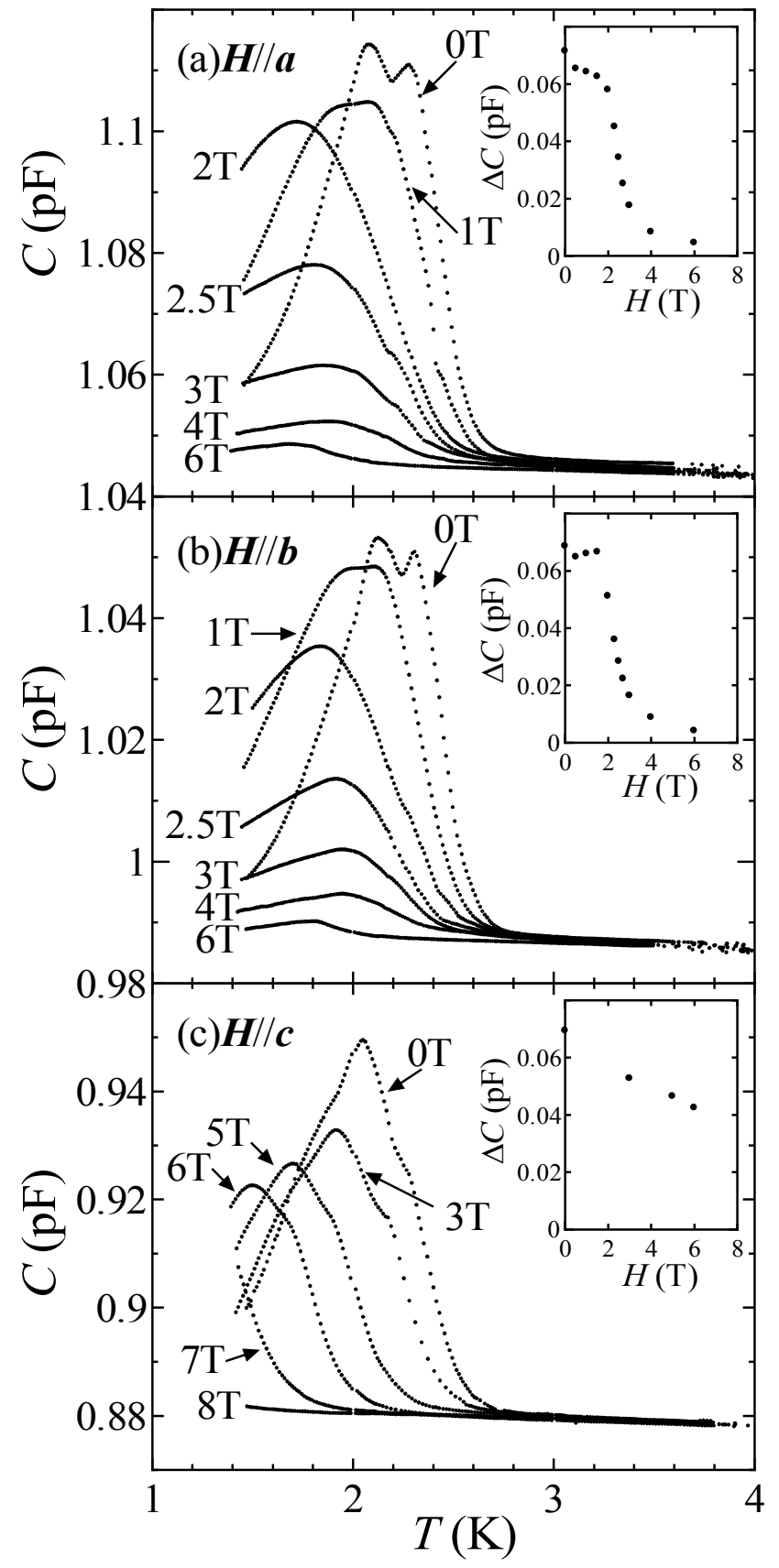

Fig. 1. $T$-dependence of the capacitance $C$ of $LiVCuO_4$ taken for the electric field $\boldsymbol{E}//\boldsymbol{a}$ under various magnetic field (a)$\boldsymbol{H}//\boldsymbol{a}$, (b)$\boldsymbol{H}//\boldsymbol{b}$ and (c)$\boldsymbol{H}//\boldsymbol{c}$. In the insets, the peak height $\Delta C$ observed at $T_N$ is shown against $H$.

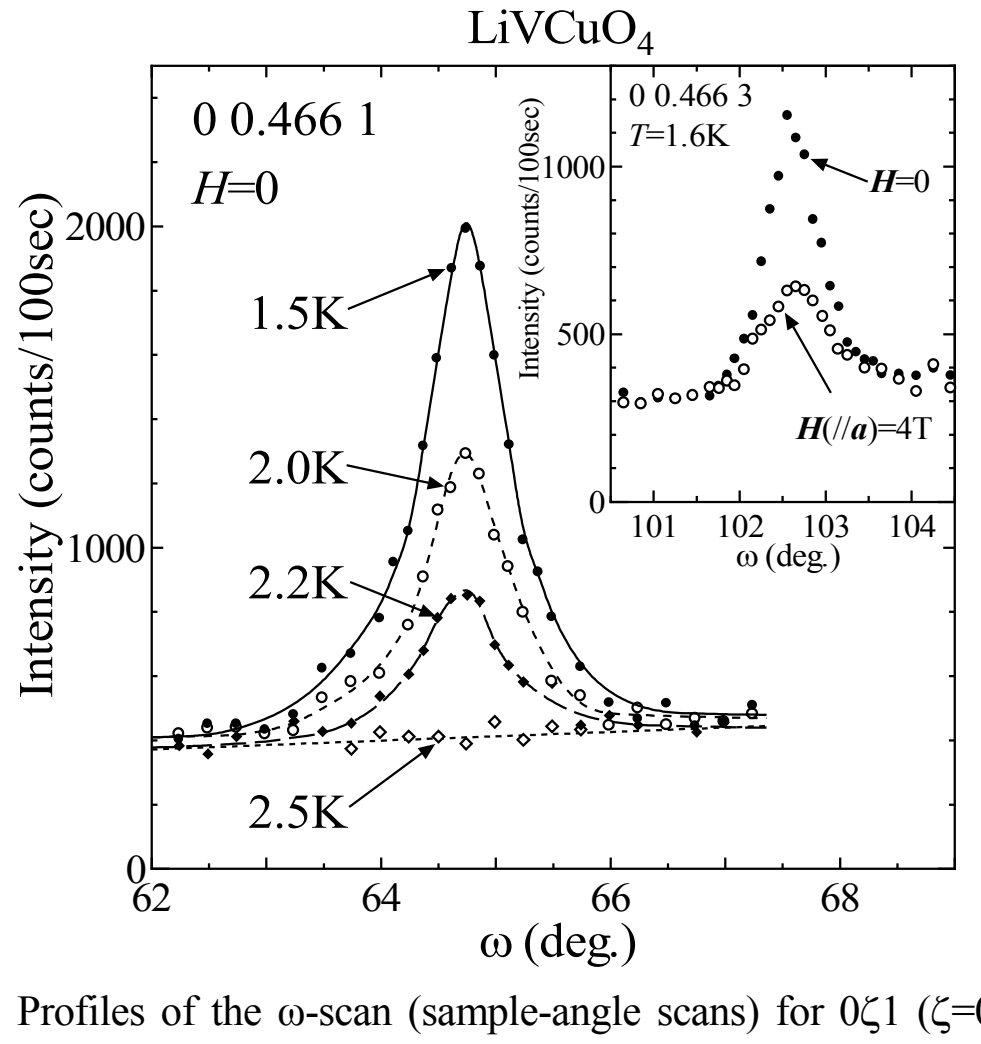

Fig. 2. Profiles of the ω-scan (sample-angle scans) for 0ζ1 (ζ=0.466) reflection taken at various fixed temperatures. The inset shows the profiles of the ω-scan for 0ζ3 reflection taken under the magnetic field $\boldsymbol{H}//\boldsymbol{a}$. The values of $H$ and $T$ are shown in the figures.

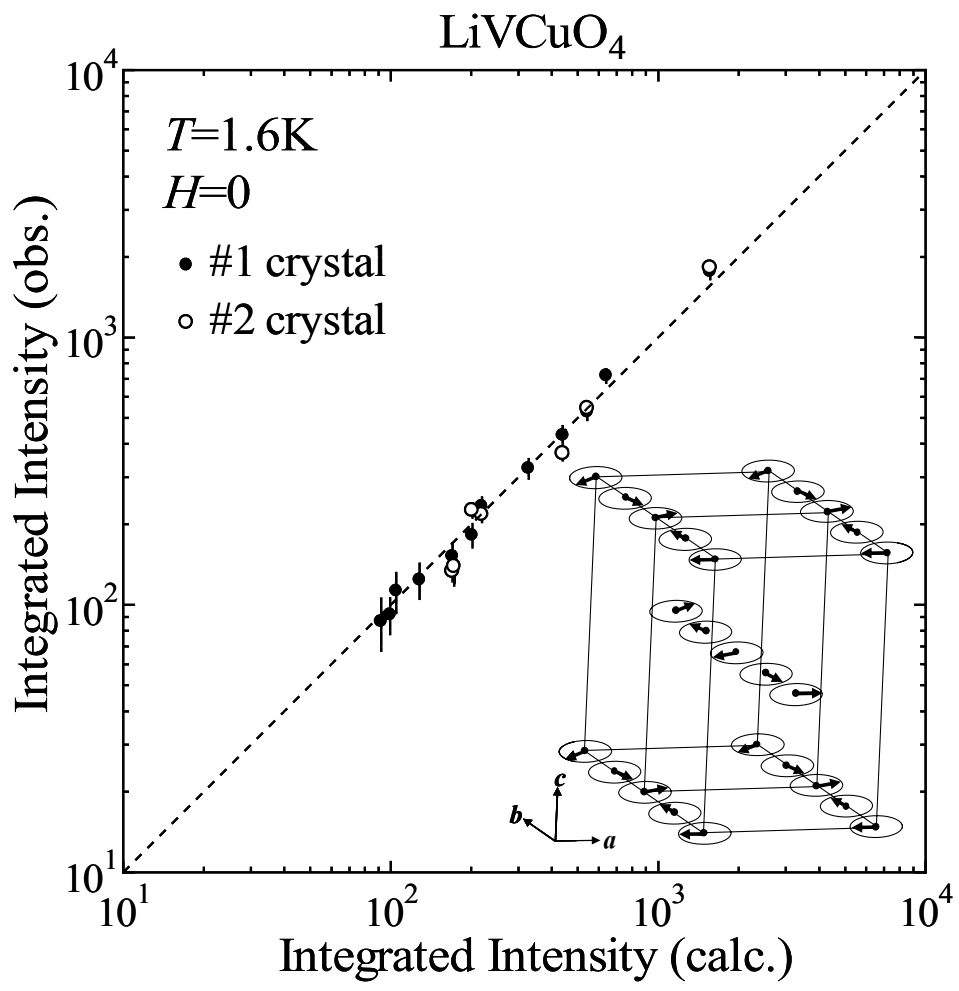

Fig. 3. Integrated intensities of the magnetic reflections of $LiVCuO_4$ observed at 1.6 K are plotted against the values calculated for the magnetic structure shown in the inset.

Next, neutron diffraction measurements have been carried out in the (0*kl*) scattering plane under the applied magnetic field

***H***(//***a***). At *H*=4 T, magnetic reflections have been observed only at the ***Q***-points where the magnetic reflections are observed at *H*=0. Examples of the observed ω-scan (sample-angle scan) profiles of the 0ζ3 reflection taken under *H*=0 T and 4 T are shown in the inset of Fig. 2. Under the field *H*=4 T along *a*, the integrated intensity is significantly suppressed and the profile width broadens. Both of the profile width observed at *H*=0 T and 4 T are larger than that of the nuclear scattering, indicating that the spatial correlation of the magnetic ordering is not ideally long ranged even at *H*=0 T. However, because the correlation length at *H*=0 T and 4 T are estimated to be rather large (~300 Å and ~200 Å, respectively), the moment system can be considered to have the quasi long range order.

In the inset of Fig. 4, we plot the ratio of the integrated intensities taken under *H*=4 T and *H*=0 T, *I*(4 T)/*I*(0 T) against 2θ for several magnetic reflections at *T*=1.6 K. The values of *I*(4 T)/*I*(0 T) of these magnetic reflections are found not to have equal values, indicating that the ordering patterns in *H*=4 T and *H*=0 T are different. Then, the magnetic structure at *H*=4 T is analyzed as follows. Because the ***Q***-points of the magnetic reflections observed at *H*=4 T are same as those at *H*=0, the periodicity of the ordering pattern is unchanged upon applying the field *H*=4 T along *a*. The observed values of *I*(4 T)/*I*(0 T) are found to be reproduced by using the helical structures with the moments within the *bc*- and *ab*-planes for *H*=4 T and 0 T, respectively, as shown in the inset of Fig. 4 by open and closed circles for the calculated and observed data, respectively. In the analyses, the ordered moment of 0.23(±0.03) $\mu_B$ is used at 1.6 K at *H*=4 T, which should be compared with the zero-field value of 0.25(±0.03) $\mu_B$.

In the *H*-dependence of the scattering intensities of 0ζ1 and 0ζ3 reflections, a clear anomalies are observed at *H*~2 T applied along *a* (not shown). These results indicate that the spin flop transition takes place at *H*~2 T, from the *ab*-plane helical to the *bc*-plane helical structure.

The main panel of Fig. 4 shows the *T*-dependence of the peak intensities for 0ζ1 reflection taken under various fixed magnetic field ***H***//***a***, together with that of the integrated intensities at *H*=0. The intensity-*T* curves below $T_N$ in the region of *H*<2 T are convex, while the curves in the region of *H*>2 T are concave, indicating that the *bc*-plane helical structure gradually grows with decreasing *T*.

The upper panel of Fig. 5 shows the *T*-dependence of the spontaneous polarization along *a*, $P_a$ taken under various magnetic field ***H***(//***a***), together with that of the neutron integrated intensities in the zero field. The saturated value of $P_a$ in *H*=0 is estimated to be ~43 μC/m$^2$, which is ten times smaller than those of the Mn-based multiferroic systems.[1,2] $P_a$ decreases with increasing *H*, and the ferroelectricity is almost entirely suppressed by the magnetic field *H*>2 T. We can see that $P_a$ is proportional to the neutron scattering intensity. In ref. 15, the appearance of the spin cholesteric state, which induces the ferroelectricity is theoretically predicted above $T_N$, when spins satisfy certain conditions. However, in $LiVCuO_4$, the ferroelectric transition is considered to occur simultaneously with magnetic phase transition within the experimental error bar as shown in Fig. 5.

No ferroelectric polarization appears with decreasing *T* along *c* within the experimental accuracy under the fields of 0 and 4 T applied along *a* in the measured *T* range as shown in the bottom panel of Fig. 5. We have observed no anomaly in the *T*-dependence of the electric susceptibility ε for ***E***//***c*** down to 1.5 K under the field *H* applied along *a* up to 8 T, either. No evidence has been found for the appearance of the ferroelectricity along ***c***, even in the region of *H* >2 T applied along *a*.

In $LiVCuO_4$, the modulation vector ***Q*** is along $\boldsymbol{b}^*$ (~0.466$\boldsymbol{b}^*$), the helical axis $\boldsymbol{e}_3$ is perpendicular to the $CuO_4$ square planes ($\boldsymbol{e}_3//\boldsymbol{c}$) at $H$=0, and $\boldsymbol{P}$ is observed along $a$, these properties indicate that the relation $\boldsymbol{P} \propto \boldsymbol{Q}\times\boldsymbol{e}_3$ holds, as was proposed by a phenomenological[11,12] and microscopic[13-16] models. According to the theories, $P$ is proportional to the $m^2$, $m$ being the magnetic order parameter. Actually as shown in Fig. 5, $P_a$ is found to be proportional to the neutron scattering intensities $I$ ($\propto m^2$), which is consistent with the above predictions.

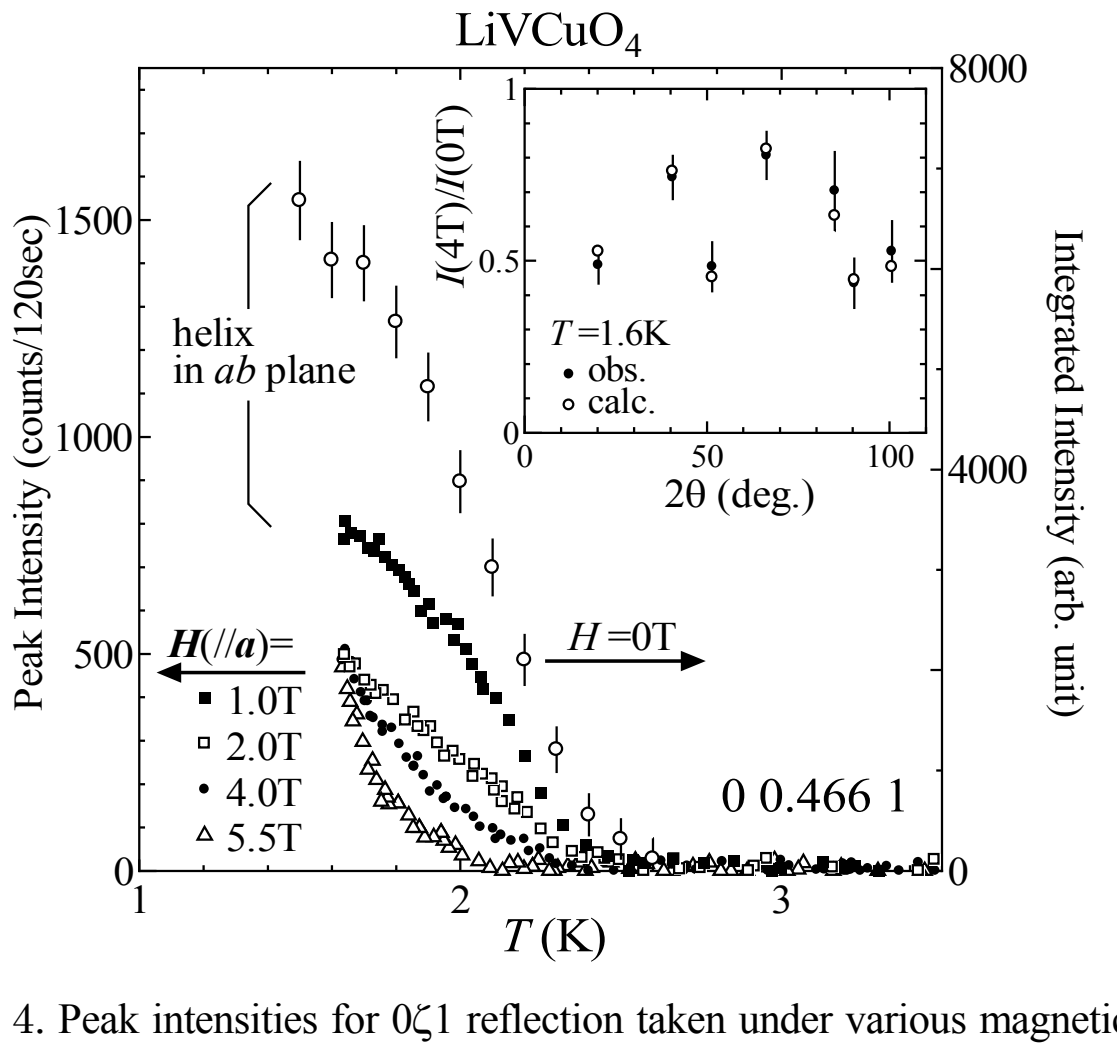

Fig. 4. Peak intensities for 0ζ1 reflection taken under various magnetic field ***H***//***a*** are shown against *T*, together with the *T*-dependence of the integrated intensities in zero field. In the inset, the ratios of the integrated intensities of the magnetic reflections taken under *H*=4 T and *H*=0 T, *I*(4 T)/*I*(0 T) are shown against 2θ at *T*=1.6 K. The open symbols show the results of the calculation obtained by using the *bc*-plane helical structure at *H*=4 T. The details of the analysis are described in the text.

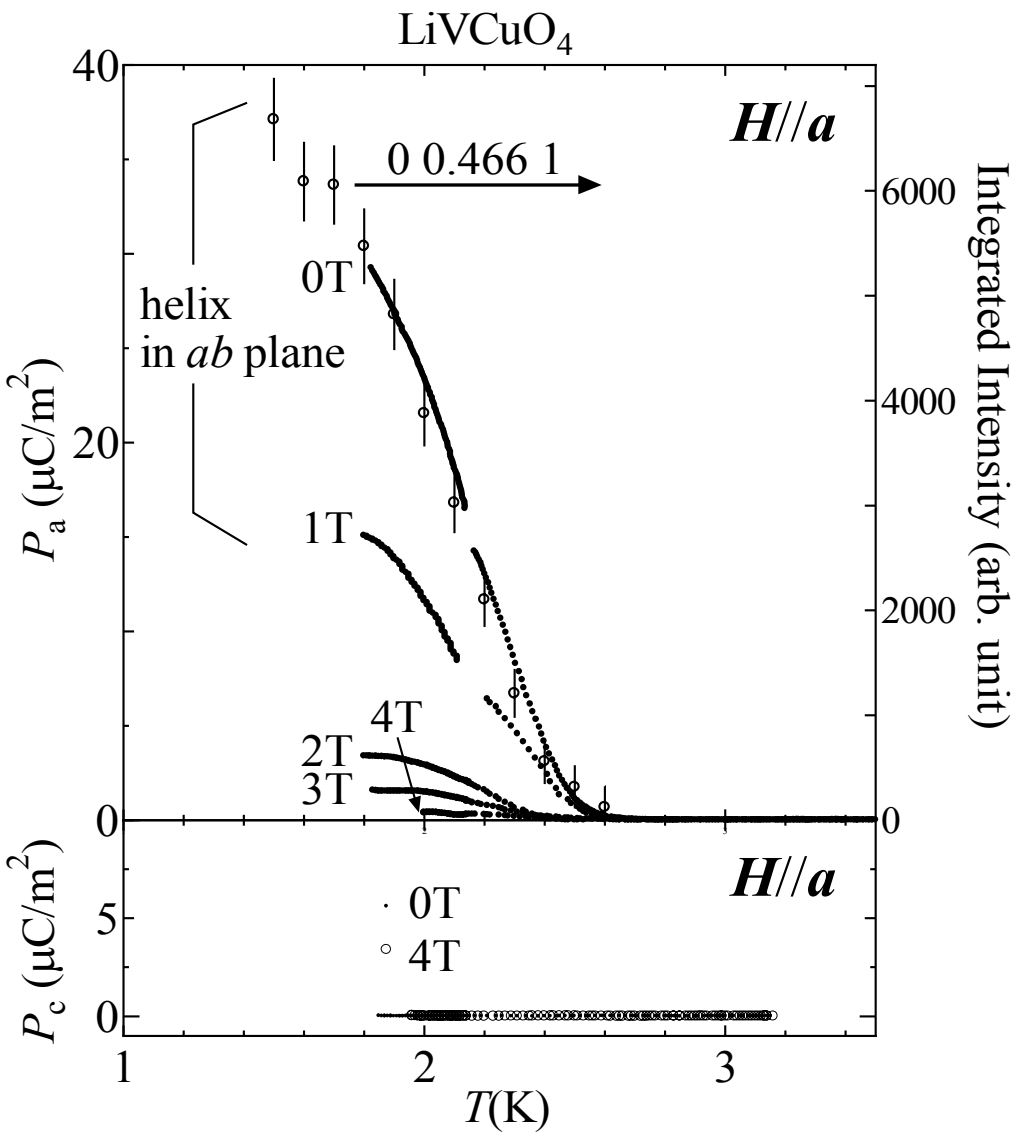

Fig. 5. Upper panel shows the $T$-dependence of the spontaneous polarization along $a$, $P_a$ of $LiVCuO_4$ taken at various fixed field $\boldsymbol{H}//\boldsymbol{a}$, together with the neutron integrated intensities for 0ζ1 reflection at $H$=0. The bottom figure shows the $T$-dependence of the spontaneous polarization $P_c$.

Xiang and Whangbo calculated the ferroelectric polarization of $LiVCuO_4$ using first principle method considering the spin-orbit coupling effect.[16] When the spin-orbit coupling is considered on both the Cu and O sites (only on the O sites) and shifts of the atomic positions are neglected (considering the asymmetric electron density distribution), they estimated $P_a$=103.5 (or 29.8) μC/m$^2$. These values seem to be consistent with the present value $P_a$~43 μC/m$^2$.

For all models stated above, the ferroelectricity is expected to be induced along $c$ or perpendicular to the $CuO_4$ square planes, when the magnetic structure is the $bc$-plane helical structure with the modulation vector along $\boldsymbol{b}^*$. However, we have not observed the ferroelectricity along $\boldsymbol{c}$ under the field $H$>2 T applied along $a$, where the $bc$-plane helical structure is actually realized. A similar result has been found in $Li_2ZrCuO_4$, which also has the edge-sharing $CuO_2$ ribbon chains and exhibits the magnetic ordering of the $bc$-plane helical structure (conventionally using the crystal axes as for the present system). For poly-crystal samples of $Li_2ZrCuO_4$, we have not observed the ferroelectricity, either.[19] Understanding of this non-observation of the ferroelectricity remains as a future problem.

In conclusion, we have shown the experimental results of the ferroelectric polarization and neutron magnetic reflections obtained under the applied magnetic field for single-crystal samples of $LiVCuO_4$. In $H$=0, the $ab$- plane helical structure reported by Gibson *et al.* [9] has been confirmed. The magnetic structure under the field along $a$ has been determined, where the spin flop transition occurs at $H$~2 T and the $ab$-plane structure changes to the $bc$-plane helical one with increasing the field through ~2 T. $P_a$ is proportional to the neutron scattering intensity at $H$=0, indicating that the magnetic order is closely related to the appearance of the electric polarization. $P_a$ decreases with increasing $H$ along $a$ and disappears in the region of $H$ > 2 T. However, we have not observed the ferroelectricity along $c$ under $H$>2 T applied along $a$, where the $bc$ plane helical structure is realized. It seems to contradict to the theoretical predictions.

**Acknowledgments**

Neutron scattering was performed within the national user's program of the Neutron Science Laboratory of the Institute for Solid State Physics (NSL-ISSP). This work has been supported by Grants-in-Aid for Scientific Research from the Japan Society for the Promotion of Science (JSPS) and by Grants-in-Aid on Priority Area from the Ministry of Education, Culture, Sports, Science and Technology.